\documentclass[twocolumn,prc,showpacs,preprintnumbers,amsmath,amssymb,superscriptaddress]{revtex4}
\usepackage{graphicx,color}
\usepackage{amssymb}
\usepackage{amsmath}
\usepackage{latexsym}
\usepackage{amsxtra}
\usepackage{amscd}
\usepackage{dcolumn}
\usepackage{bm}

\graphicspath{}
\let\oldsqrt\sqrt
\def\sqrt{\mathpalette\DHLhksqrt}
\def\DHLhksqrt#1#2{\setbox0=\hbox{$#1\oldsqrt{#2\,}$}\dimen0=\ht0
\advance\dimen0-0.2\ht0
\setbox2=\hbox{\vrule height\ht0 depth -\dimen0}%
{\box0\lower0.4pt\box2}}

\usepackage{hyperref}

\begin{document}

\title{Elastic breakup cross sections of well-bound nucleons}

\author{K.~Wimmer}
\affiliation{Department of Physics, Central Michigan University, Mt. Pleasant, Michigan 48859, USA}
\affiliation{National Superconducting Cyclotron Laboratory, Michigan State University, East Lansing, Michigan 48824, USA}

\author{D.~Bazin}
\affiliation{National Superconducting Cyclotron Laboratory, Michigan State University, East Lansing, Michigan 48824, USA}

\author{A.~Gade}
\affiliation{National Superconducting Cyclotron Laboratory, Michigan State University, East Lansing, Michigan 48824, USA}
\affiliation{Department of Physics and Astronomy, Michigan State University, East Lansing, Michigan 48824, USA}

\author{J.~A.~Tostevin}
\affiliation{National Superconducting Cyclotron Laboratory, Michigan State University, East Lansing, Michigan 48824, USA}
\affiliation{Department of Physics, Faculty of Engineering and
Physical Sciences, University of Surrey,
Guildford, Surrey GU2 7XH, United Kingdom}
\author{T.~Baugher}
\affiliation{National Superconducting Cyclotron Laboratory, Michigan State University, East Lansing, Michigan 48824, USA}
\affiliation{Department of Physics and Astronomy, Michigan State University, East Lansing, Michigan 48824, USA}

\author{Z.~Chajecki}
\affiliation{National Superconducting Cyclotron Laboratory, Michigan State University, East Lansing, Michigan 48824, USA}

\author{D.~Coupland}
\affiliation{National Superconducting Cyclotron Laboratory, Michigan State University, East Lansing, Michigan 48824, USA}
\affiliation{Department of Physics and Astronomy, Michigan State University, East Lansing, Michigan 48824, USA}

\author{M.~A.~Famiano}
\affiliation{Department of Physics, Western Michigan University, Kalamazoo, Michigan 49008, USA}

\author{T.~K.~Ghosh}
\affiliation{Variable Energy Cyclotron Centre, 1/AF Bidhannagar, Kolkata 700064, India}

\author{G.~F.~Grinyer}\affiliation{Grand
        Acc\'el\'erateur National d'Ions Lourds (GANIL),
        CEA/DSM-CNRS/IN2P3, Bvd Henri Becquerel, 14076 Caen, France}

\author{M.~E.~Howard}
\affiliation{Department of Physics and Astronomy, Rutgers University, New Brunswick, New Jersey 08903, USA}

\author{M.~Kilburn}
\author{W.~G.~Lynch}
\affiliation{National Superconducting Cyclotron Laboratory, Michigan State University, East Lansing, Michigan 48824, USA}
\affiliation{Department of Physics and Astronomy, Michigan State University, East Lansing, Michigan 48824, USA}

\author{B.~Manning}
\affiliation{Department of Physics and Astronomy, Rutgers University, New Brunswick, New Jersey 08903, USA}

\author{K.~Meierbachtol}
\affiliation{National Superconducting Cyclotron Laboratory, Michigan State University, East Lansing, Michigan 48824, USA}
\affiliation{Department of Chemistry, Michigan State University, East Lansing, Michigan 48824, USA}

\author{P.~Quarterman}
\author{A.~Ratkiewicz}
\author{A.~Sanetullaev}
\author{R.~H.~Showalter}
\author{S.~R.~Stroberg}
\affiliation{National Superconducting Cyclotron Laboratory, Michigan State University, East Lansing, Michigan 48824, USA}
\affiliation{Department of Physics and Astronomy, Michigan State University, East Lansing, Michigan 48824, USA}

\author{M.~B.~Tsang}
\affiliation{National Superconducting Cyclotron Laboratory, Michigan State University, East Lansing, Michigan 48824, USA}

\author{D.~Weisshaar}
\affiliation{National Superconducting Cyclotron Laboratory, Michigan State University, East Lansing, Michigan 48824, USA}

\author{J.~Winkelbauer}
\affiliation{National Superconducting Cyclotron Laboratory, Michigan State University, East Lansing, Michigan 48824, USA}
\affiliation{Department of Physics and Astronomy, Michigan State University, East Lansing, Michigan 48824, USA}

\author{R.~Winkler}
\affiliation{National Superconducting Cyclotron Laboratory, Michigan State University, East Lansing, Michigan 48824, USA}

\author{M.~Youngs}
\affiliation{National Superconducting Cyclotron Laboratory, Michigan State University, East Lansing, Michigan 48824, USA}
\affiliation{Department of Physics and Astronomy, Michigan State University, East Lansing, Michigan 48824, USA}

\begin{abstract}
The $^{9}$Be($^{28}$Mg,$^{27}$Na) one-proton removal reaction with a large proton separation
energy of $S_\text{p}(^{28}$Mg)=16.79 MeV is studied at intermediate beam energy. Coincidences of
the bound $^{27}$Na residues with protons and other light charged particles are measured. These
data are analyzed to determine the percentage contributions to the proton removal cross section from
the elastic and inelastic nucleon removal mechanisms. These deduced contributions are compared with
the eikonal reaction model predictions and with the previously measured data for reactions
involving the removal of more weakly-bound protons from lighter nuclei. The role of transitions
of the proton between different bound single-particle configurations upon the elastic breakup
cross section is also quantified in this well-bound case. The measured and calculated elastic
breakup fractions are found to be in good agreement.
\end{abstract}

\date{\today}
\maketitle

\section{Introduction}

Nucleon removal reactions are a very effective means to both populate nuclei
far from stability with relatively high yields and to probe their structure.
The use of projectile beams of high energy allows thick reaction targets to be
used providing sufficient luminosity for precise measurements, even for very
exotic systems. These high incident energies (few 100 MeV per nucleon) also allow
certain simplifications in the theoretical description of the reaction dynamics,
namely use of the sudden (fast collision) and eikonal (forward scattering)
approximations. In general, three physical mechanisms may contribute to the
one-nucleon removal reaction cross section. Their relative importance depends
sensitively on the mass and charge of the target nucleus and the separation
energy of the removed nucleon from the projectile ground-state. In the case
of very weakly-bound nucleons and a heavy, highly-charged target nucleus the
reaction is often dominated by elastic Coulomb breakup and large soft-$E1$
excitation strength to low relative energy two-body states of the (residue +
nucleon) break-up continuum; see for example \cite{nakamura99,nakamura13} and
references therein.

For light target nuclei, most often $^9$Be and $^{12}$C, and the removal of a
more well-bound nucleon, our primary interest here, such Coulomb dissociation
contributions are negligible and the reaction proceeds by the strong interaction.
The two contributing mechanisms are then: (i) elastic breakup of the projectile, also
called diffraction dissociation, where the differential forces acting between the
constituents and the target dissociate the projectile but leave the target nucleus
in its ground state, and (ii) inelastic breakup, where the interactions that remove the
nucleon transfer energy to and remove the target nucleus from the elastic channel.
This second mechanism is often called stripping. Since these two reaction mechanisms
lead to distinct final states, their cross sections can be added and, since most
intermediate energy experiments measure only the projectile-like residues (after
nucleon removal), these are normally compared with the sum of model calculations of
the cross sections computed due to the two mechanisms. However, given the now significant
body of experimental data that shows systematic differences of these measured
removal yields from those calculated using shell-model plus reaction theory inputs,
shown in Refs. \cite{gade08,tostevin12}, particularly those involving well-bound
nucleons, the quality of the theoretical predictions of these two mechanisms is
significant to validate the models used. We note that in the case of a $^9$Be
target, that is itself weakly bound with $S_\text{n}$=1.66 MeV and with no bound
excited states, the stripping mechanism at energies near 100 MeV per nucleon will
in general populate a many-particle final state and several light fragments,
including the removed nucleon.

There have been extensive investigations of few-body models for the elastic
breakup of the deuteron and well-clustered nuclei due to the Coulomb and nuclear
interactions. Many have been extended, refined and applied for the study of light
weakly-bound (halo) nuclei. These, in general, provide an excellent description of
the increasing body of available elastic scattering and elastic breakup data on
both one- and two-neutron halo systems. To account for the many-body nature of light
halo nuclei, more microscopic and ab-initio structure/reaction treatments are also
being developed. Few-body methods now include Faddeev-based, coupled-channels and
multiple-scattering
quantum methods, eikonal-like and other semi-classical approaches, each optimal
for different ranges of projectile energies and target nuclei but with valuable
regions of overlap. In contrast, the reaction dynamics of elastic (and inelastic)
breakup of more bound nucleons has received relatively little recent theoretical
or experimental attention. Here, published calculations have been restricted to
the use of eikonal models (e.g. \cite{tostevin01b,hansen03,gade08} and multiple references
therein) and the transfer to the continuum (TC) technique \cite{angela88,flavigny12};
although the latter is not applicable to proton removal and in its usual implementation
uses approximations that are also poorly-suited to reactions involving well-bound
neutrons \cite{angela01,alvaro05}.

Early experimental studies of the removal reaction mechanism~\cite{zinser97} also
focused on the study of neutron removal from light neutron halo nuclei. In these
experiments, neutron detectors covered only very forward angles and the elastic
breakup component of the cross section was measured exclusively. Based on a quite
simple geometrical model and the assumed dominance of the asymptotic region of
the neutron wave function, it was suggested that the relative contribution to the
removal reaction cross section from the diffractive mechanism will
decrease with the neutron separation energy as $1/\sqrt{S_\text{n}}$ -- a relatively rapid fall in the
diffraction component with $S_\text{n,p}$. Intermediate-energy eikonal model
calculations for both well- and weakly-bound nucleons on the other hand suggest that
the single-particle removal cross sections (stripping plus diffraction) are strongly
correlated with the root mean squared (rms) radius of the orbital from which the
nucleon is removed, see e.g. Fig.\ 2 of Ref. \cite{gade08}. The calculations also
suggest a weaker dependence of the elastic breakup contribution on separation energy
and that this component persists and contributes significantly to the removal of
well-bound nucleons. This paper aims to quantify this fractional elastic breakup contribution
using more exclusive measurements for the $^{9}$Be($^{28}$Mg,$^{27}$Na) one-proton
removal reaction with proton separation energy $S_\text{p}(^{28}$Mg) = 16.79~MeV.

The first precise measurement of the individual contributions from the stripping and
diffraction mechanisms to nucleon removal~\cite{bazin09} also exploited weakly-bound
projectiles, $^8$B and $^9$C, having proton separation energies of $S_\text{p}$=
0.137 MeV and 1.296 MeV, respectively. The measured contributions from the diffractive
and stripping mechanisms were found to be in good agreement with the predictions using
the eikonal reaction dynamics description. That analysis also made use of continuum
discretized coupled channels (CDCC) calculations of the elastic breakup to correct the
measurements for cross section that was unobserved due to the restricted solid angle
coverage for proton and light charged particle detection in the experiment.

In this work, we extend this earlier study of exclusive reaction mechanism measurements
to the removal of strongly-bound protons. The data are compared with eikonal reaction
model calculations and the role of transitions of the proton between bound single-particle
configurations is quantified in this strongly-bound case.  We show conclusively that the
importance of the elastic breakup mechanism is not limited to loosely-bound systems but
persists in the removal of well-bound nucleons. The dependence of the elastic breakup
fraction on $S_\text{p}$ will be discussed.

We study the $^{9}$Be($^{28}$Mg,$^{27}$Na) reaction that is well suited as a test case
since: (i) the proton separation energy from $^{28}$Mg is large ($S_\text{p} = 16.79$
MeV), and (ii) intense intermediate-energy beams of $^{28}$Mg are available. Furthermore,
since both $^{28}$Mg and its reaction residue $^{27}$Na have a simple structure, due to
the $N=16$ sub-shell closure, their spectra are very well described by shell-model
calculations in the $sd$-shell model space. Based on shell-model spectroscopy, the
one-proton removal reaction from $^{28}$Mg is expected to mainly populate the $^{27}$Na
ground state. No $\gamma$-ray detection was used to identify the $^{27}$Na final state
in the present measurement. The new experimental data are compared with the earlier
data for the loosely bound $^8$B and $^9$C systems~\cite{bazin09}.

\section{Experimental setup}

The $^{28}$Mg beam was produced by fragmentation of an $^{40}$Ar primary beam with an
energy of 140~MeV/u, provided by the coupled cyclotron facility at the National
Superconducting Cyclotron Laboratory (NSCL), on a 846~mg/cm$^2$ $^{9}$Be production target.
The desired fragment with an energy of 93~MeV/u was selected with the A1900 fragment
separator~\cite{morrissey03} and impinged on a 9~mg/cm$^2$ $^{9}$Be secondary target at
the target position of the S800 high resolution magnetic spectrograph~\cite{bazins80003}.
Incoming beam particles were identified event-by-event by their time-of-flight between
two plastic scintillators before the target. The average rate on target was $5\cdot10^5$
$^{28}$Mg/s. The largest contaminant, $^{29}$Al, amounted to only 1.5\% of the total
beam intensity.

The reaction residues were identified by an energy loss measurement in an ionization chamber
in the focal plane detector box of the S800 spectrograph and the time-of-flight between
a scintillator before the target and one in the focal plane. Momentum and energy of the
reaction residues were reconstructed from the magnetic rigidity setting of the S800
spectrograph and the angles and positions of particles in the focal plane, measured with
position sensitive cathode readout drift chambers (CRDC). Light charged particles from
the removal reaction were detected in coincidence with the heavy $^{27}$Na residue in
the high resolution array HiRA~\cite{wallace07}. $\Delta E-E$ telescopes based on 1.5~mm
thick double-sided silicon strip detectors (DSSSD) and 4~cm long CsI crystals allowed
for unambiguous event-by-event identification of light charged particles, such as protons,
deuterons, tritons, $^3$He and $\alpha$ particles. The HiRA array covered polar angles
($\vartheta$) from 9$^\circ$ to 56$^\circ$ and for a given value of $\vartheta$ up to
40\% of the azimuthal angles were covered. A correction for the limited azimuthal
acceptance within the $\vartheta$ range was applied following our previous work~\cite{wimmer12}.

Two settings of the S800 spectrograph were used. Both optical settings have distinct advantages
which are described below. The first optics mode, (i), is the so-called focused mode. The
second mode, (ii), the so-called dispersion matched mode, offers a higher resolution.
Mode (i): In order to precisely measure the inclusive one-proton removal reaction cross section,
the beam was focused on the target. Corrections for the incoming momentum dispersion were made
by measuring the position and angle at the intermediate image before the target with two position
sensitive parallel plate avalanche counters (PPAC). In the focused mode, acceptance cuts were
limited and well under control. Since the reaction point on the target is well defined the
angle of the light charged particles could be determined with high precision. However, the
momentum resolution for the heavy residues is then limited to $\Delta p = 24$~MeV/c.
Mode (ii): In contrast to the high-resolution mode, the spectrograph was operated in the dispersion
matched mode allowing for a precise measurement of the momentum change in the reaction
($\Delta p = 4.6$~MeV/c). The disadvantage of this method, since the dispersion of the beam
at the target plane is very large, is that only part of the incoming beam hits the target.
Thus, a measurement of the absolute reaction cross section cannot be obtained and only relative
values are reported. A second, additional uncertainty in this measurement mode is the uncertainty
in the event-by-event position of the reaction (in the dispersive direction) in the target plane.
Since the beam is spread over the target, the emission angle of the light particle can only
be obtained with a $\vartheta$ uncertainty of $6-9^\circ$, depending on the angle.

\section{Experimental data analysis \label{exp_anal}}

The inclusive one-proton removal reaction cross section was measured in the focused mode.
No coincidence with particles detected in HiRA was required. Due to the finite acceptance
of the S800 spectrograph, corrections at the largest and smallest residue momenta had to
be applied. In total, these corrections amount to 2\% of the inclusive cross section.
For the $^{9}$Be($^{28}$Mg,$^{27}$Na) reaction an inclusive cross section of 36(1)~mb
was obtained.

\begin{figure}[ht]
\centering
\includegraphics[width=\columnwidth]{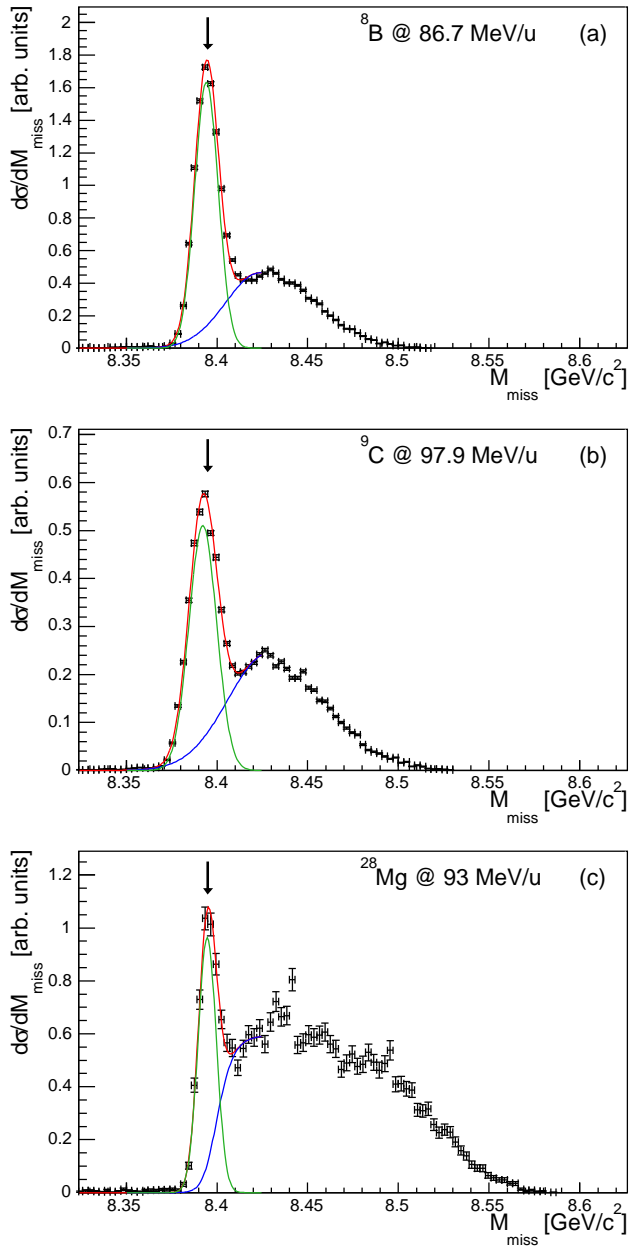}
\caption{(Color online) Missing mass spectra of the one-proton removal reaction
residues $^7$Be (a), $^8$B (b) and $^{27}$Na (c) and a proton
detected in coincidence. The beam energies are shown in each panel. The arrows
show the position of the target mass,
$M(^{9}\text{Be})$, and indicate events associated with proton removal by the
elastic breakup mechanism. The $^{28}$Mg spectrum is from the current measurement
and the $^8$B and $^9$C spectra were extracted from the data sets of Ref.\
\cite{bazin09}.}
\label{fig:missmass}
\end{figure}

In coincidence with the $^{27}$Na residue, light charged particles are detected
in HiRA. Events with deuterons, tritons or heavier particles are associated with
the inelastic removal (stripping) mechanism since the additional nucleons can only
originate from the target nucleus. Both elastic and inelastic removal processes
contribute to events where $^{27}$Na nuclei and a proton are detected in coincidence.
In elastic breakup events, the energy of the detected proton and the corresponding
one-proton removal residue are correlated. This correlation can be investigated
through the missing mass of the event which is reconstructed from the momenta and
energies (the momentum four-vectors $P$) of the $^{27}$Na residue and
the proton.
\begin{equation*}
M_\text{miss} = \sqrt{(P_\text{beam}+M_\text{target} - P_\text{p} - P_\text{Na})^2}
\end{equation*}
This missing mass spectrum is shown in the lower panel in Fig. \ref{fig:missmass}.
Figure \ref{fig:missmass} also presents the missing mass spectra for the previously
measured cases, weakly-bound proton removal from the light nuclei $^8$B and $^9$C
\cite{bazin09} for comparison with the present case. The elastic breakup events are
characterized by the peak in the
missing mass corresponding to the mass of the target nucleus, $M(^{9}\text{Be})=
8.395$~GeV/c$^2$. In the elastic breakup mechanism the nucleon is removed in a
collision that leaves the target nucleus in its ground state. As no energy is lost
to the target nucleus in this case, this leads to a sharp peak in the missing mass. In
stripping events on the other hand, where there is energy transfer and the target
is excited to a greater or lesser degree, the missing mass for such events is
higher than the target mass. The continuous broad distributions seen in all
panels in Fig.~\ref{fig:missmass} are thus attributed to inelastic breakup
events. We note that the larger proton separation energy of the present data
set results in the relative contribution from elastic breakup to these proton
coincidence events being significantly reduced. This is to be expected for the
proton separation energy of $S_\text{p} = 16.79$~MeV, as compared to $S_\text{p}
= 1.296$~MeV and 0.137~MeV for the $^9$C and $^8$B systems, respectively.

In order to quantify the elastic breakup component to the removal cross section
this missing mass spectrum is fitted. In the case of
  the weakly bound $^9$C and $^8$B systems we show a function
  consisting of two Gaussians following the earlier work of~\cite{bazin09}. Numerical values for the relative amount of the
  elastic breakup cross section discussed later are taken from~\cite{bazin09}.
The fitting function in the lower panel consists of a Gaussian,
to describe the diffraction peak, and a smooth sigmoid step function, to estimate
the background from stripping events below
$M_\text{miss}=8.41$~GeV/c$^2$. 
\begin{eqnarray*}
f(M_\text{miss}) = \frac{N}{\sqrt{2\pi}\sigma}\exp{\left[-\left(
      \frac{M_\text{miss}-M_0}{\sqrt{2}\sigma}\right)^2\right]}\nonumber \\
+ N^\prime  \left[1+\exp\left(\frac{M^\prime_0-M_\text{miss}}{\sigma^\prime}\right)\right]^{-2}
\end{eqnarray*}
The choice for this shape is motivated by the
  assumption that the elastic breakup leads to a sharp peak (delta function) at
  $M_\text{miss} = M(^{9}\text{Be})$ and the inelastic component only
  contributes for $M_\text{miss} > M(^{9}\text{Be})$. These ideal
  shapes are then smeared with the experimental resolution. 
All six parameters of the fit function are left free to
  vary. Within the uncertainty of the resulting values we find $M_0 =
  M^\prime_0$ and $\sigma = \sigma^\prime$.
The width of the elastic peak $\sigma$ is in agreement
  with the expected resolution given by the momentum width of the incoming beam,
the differential energy loss in the target, and the energy and angle resolution
for protons detected in HiRA. It is more narrow in the case of the proton removal
reaction from $^{28}$Mg because of the different optics mode used for the S800.
The data for proton removal from $^{28}$Mg shown in Fig.~\ref{fig:missmass}
(lower panel) were recorded in dispersion matched mode, while the data sets for
the weakly-bound light nuclei (upper and central panels) were measured in
focused mode -- in order to extract the proton angular
distribution~\cite{bazin09}.
In order to extract the elastic contribution to the
  breakup cross section from Fig.~\ref{fig:missmass} the part of the inelastic
contribution leaking into the elastic peak has to be obtained from the
fitting procedure. This is the major source of uncertainty
in the extraction of the elastic cross section. We have estimated this
systematic uncertainty to be $5$~\% by varying the shape of the sigmoid function, and by
fixing certain parameters in the fit. 
 Before the cross section for the elastic break
up process can be compared to theoretical predictions, additional
systematic corrections for misidentification of high and low energy protons
must be applied because Fig.~\ref{fig:missmass} only contains events
with proton kinetic energies between 15 and 120~MeV.

The measured cross section for $^{27}$Na and proton coincidences has to be corrected
for the misidentification of particles with very low energies. Protons with kinetic
energies below 15~MeV are stopped in the silicon detectors of HiRA and cannot be
unambiguously identified. The differential cross sections as a function of the light
particle energy therefore shows a step at the energy that is required to punch
through the silicon detector. This energy is different for the different light
particles. This specific energy loss allows us to extract the number of protons
amongst the unidentified particles and correct the elastic break up cross section
accordingly. At the highest energies ($E_\text{p}\gtrsim 120$~MeV) protons will
punch through the silicon as well as the CsI detectors of HiRA with the result
that their total kinetic energy is unknown, and they cannot be unambiguously
identified using the $\Delta E-E$ technique. In order to determine the systematic
correction for misidentification of these high energy protons, the differential
cross section as a function of proton kinetic energy is analyzed for the
identified protons. This distribution shows a sharp cut-off at $E_\text{p}\approx
120$~MeV which allows us to estimate that 5\% of the elastic breakup events
result from proton kinetic energies greater than 120~MeV.

After application of these corrections, the elastic breakup fraction of the total
one-proton removal cross section can be determined relative to the inclusive
cross section for production of $^{27}$Na, discussed above. The value obtained
from the $^{27}$Na data set taken in dispersion matched mode (extracted from
Fig.~\ref{fig:missmass} for identified protons and from a similar plot for
particles with kinetic energies below the identification threshold) is 11.2(12)\%
after the aforementioned corrections have been applied. The broad distribution
in the missing-mass distribution of Fig.~\ref{fig:missmass}, for identified protons,
represents 37(2)\% of the detected inelastic breakup (stripping) events. While the
resolution in the missing mass spectrum is worse when using the data set taken
in focused mode, an elastic breakup fraction of 12.5(17)\%  was determined.
This shows that even though the two settings have distinct disadvantages, a
consistent result is obtained from the two data sets. Finally, these experimental
values need to be corrected for the missing angular acceptance of the HiRA array
for polar angles $\vartheta<9^\circ$. As in \cite{bazin09}, this correction is
estimated using continuum discretized coupled channels (CDCC) calculations. This
will be discussed in Section \ref{cdcc}.

\section{Theoretical analysis}

\subsection{Shell-model input}

As stated, the measurements made are inclusive with respect to all bound final
states of $^{27}$Na, whose ground state has $S_\text{n} = 6.73$ MeV. Shell-model
calculations using the USD $sd$-shell effective interaction \cite{wildenthal84}
are used to compute the $^{28}$Mg ground-state and the $^{27}$Na final states
and one-proton removal spectroscopic factors ($C^2S$). This shell model
proton-removal strength is predominantly to just three final states, with observed
counterparts; the $5/2^+$ ground state, with $C^2S$ = 3.137, an almost degenerate $3/2^+$
(14 keV) state with $C^2S$ = 0.124, and a $1/2^+$ (1630 keV) state with $C^2S$ =
0.304. Additional small fragments of $5/2^+$, $3/2^+$ and $1/2^+$ strength are
distributed over many excited states below the 6.73 MeV first neutron threshold.
Thus, as stated earlier, the ground-state to ground-state removal is expected to be
the dominant transition. However, given the inclusive nature of the measurements,
in the following calculations, when comparisons are made with the data, we sum
these small fragments of strength into the spectroscopic factors used for the three
states above. These become $C^2S(5/2^+)$ = 3.289, $C^2S(3/2^+)$ = 0.262 and $C^2S
(1/2^+)$ = 0.330 and are the values shown in Table \ref{tab:tab1}. Their sum, of
3.88, effectively exhausts the spectroscopic sum-rule. Given the large
ground-state to ground-state $S_\text{p}$ of 16.79 MeV, assigning the same (smaller)
excitation energy to these multiple small fragments will overestimate slightly their
cross section contribution, but this has a very small effect on the inclusive cross
section and negligible effect on the fraction of cross section due to elastic
breakup.

\subsection{Eikonal model calculations\label{eik_calcs}}

This shell-model structure information is now used with the eikonal dynamical
model \cite{tostevin01b,hansen03} to make theoretical predictions for the inclusive
one-proton removal cross section at 93 MeV per nucleon and for its components
from the elastic and inelastic breakup mechanisms. More specifically, we follow
precisely: (i) the framework for the construction of the proton- and residue-target
optical potentials and their eikonal S-matrices, from the assumed $^9$Be target and
$^{27}$Na Hartree-Fock (HF) densities, and (ii) constraints on the geometries of the bound
state radial overlap functions of the proton using analogous and consistent HF calculations
for $^{28}$Mg, as are described in detail in Ref. \cite{gade08}. All bound proton
wave functions are calculated in Woods-Saxon wells with diffuseness $a_0$ = 0.7 fm and
a spin-orbit potential strength of $V_\text{so}=6.0$ MeV. The reduced radius parameters
$r_0$ consistent with the HF constraints are 1.285, 1.322 and 1.205 fm for the $1d_{5/2}$,
$1d_{3/2}$ and $2s_{1/2}$ orbitals, respectively. The effective proton separation energies
used in comparing with the data are determined from the empirical $S_\text{p}$ and the
shell-model excitation energies in Table \ref{tab:tab1}.

\begin{figure}[ht]
\centering
\includegraphics[width=\columnwidth]{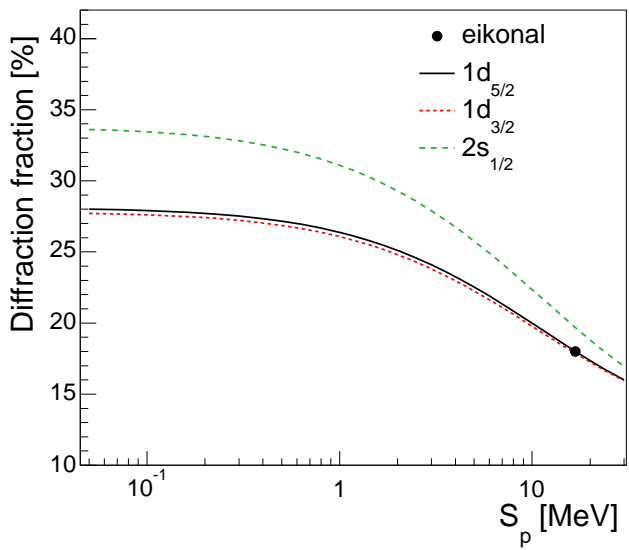}
\caption{(Color online) Computed fractional contribution (\%) of the elastic breakup
(diffraction) reaction mechanism to the proton removal cross section as a function
of the proton separation energy. The lines show the predictions of the
eikonal model for proton removal from $1d_{5/2}$ (solid black), $1d_{3/2}$ (dotted
red) and $2s_{1/2}$ (dashed green) orbitals. The theoretical prediction, 18\%,
when using spectroscopic factors calculated from the shell model (see text and
Table \ref{tab:tab1}), is shown by the filled black circle.}
\label{fig:calc_percent}
\end{figure}

Based on these reaction inputs, Fig.~\ref{fig:calc_percent} shows the calculated dependence
of the percentage fractional contribution of the elastic breakup (diffraction) mechanism
to the proton removal cross section for pure $1d_{5/2}$, $1d_{3/2}$ and $2s_{1/2}$
single-proton orbitals as a function of the proton separation energy -- from
$S_\text{p}$ = 0.05 through 20 MeV. The actual contributions from each shell-model final
state, weighted by their spectroscopic factors, at the physical $S_\text{p}$ are shown
in Table \ref{tab:tab1}.
The theoretical cross sections $\sigma_\text{th}$ include the  $[A/(A-1)]^N$ center-of-mass
correction factor to the shell-model spectroscopic factors \cite{Dieperink74}, with $N=2$
for these $sd$-shell orbitals. We see that, from these eikonal model calculations, the
computed theoretical elastic breakup fraction is 18\%. In Section
\ref{projoff} we will investigate minor corrections to these presented conventional
eikonal calculations -- due to some flux leading to transitions of the proton between
bound states and not to the breakup continuum. First we discuss the measured and calculated
inclusive cross section values and also comment on the separation energy dependence of
the calculated elastic breakup fractions as are shown by the lines in
Fig.~\ref{fig:calc_percent}.

\begin{table}[ht]
\caption{\label{tab:tab1} Shell-model states of $^{27}$Na and their spectroscopic
factors (see text) for one-proton removal from $^{28}$Mg from the USD effective
interaction \cite{wildenthal84}. We show the theoretical proton-removal cross
sections to each final state (for 93 MeV/nucleon projectiles on a $^9$Be target),
their elastic and inelastic breakup components, and the overall predicted fraction
(\%) of elastic breakup events. The predicted
summed $C^2 S$ of 3.88 to these bound configurations essentially exhausts the
maximum available strength.}
\begin{tabular}{ccccccc}
\hline \hline
\ \ $J^\pi$\ \ &$E_{\rm x}$&   $\ \ C^2S\ \ $    &\ $\sigma_{\rm th}^{\rm inel}$
\ \ &\ $\sigma_{\rm th}^{\rm elas}$&\ $\sigma_{\rm th}$& \ \ elas\ \ \\
        &(MeV)  &  & (mb)   & (mb) & (mb)& (\%)\\
\hline
 $5/2^+$    &0.000&3.289       &38.75 & 8.47 &47.22& 17.9\\
 $3/2^+$    &0.014&0.262       &2.95  & 0.64 & 3.59& 17.8\\
 $1/2^+$    &1.630&0.330       &3.86  & 0.92 & 4.78& 19.3\\
\hline
Sum &            &3.88         &45.56 & 10.03&55.59& 18.0\\
\hline\hline
\end{tabular}
\end{table}

\subsubsection{Inclusive cross section}

Regarding comparison with the measured inclusive cross section of 36(1)~mb, the
calculated cross section is seen to be 55.6 mb. The asymmetry of the neutron and
proton Fermi surfaces in $^{28}$Mg, based on their separation energies and the
dominance of the ground-state transition, means (in the notation of \cite{gade08})
that $\Delta S$ = 8.29 MeV for one-proton removal. The cross sections ratio $R_s
= \sigma_\text{exp}/\sigma_\text{th}$ is therefore 0.65(2). This value is consistent with
the systematics of this cross section ratio with $\Delta S$, referred to in the
introduction and shown in the figures of Refs. \cite{gade08,tostevin12}.

\subsubsection{Separation energy dependence\label{sepen}}

In the previously studied weakly-bound proton cases, $^9$C and $^8$B~\cite{bazin09},
of order one third of the proton removal cross section was due to elastic breakup.
The calculated elastic breakup fractions for removal from the active $1d_{5/2}$,
$1d_{3/2}$ and $2s_{1/2}$ $sd$-shell proton orbitals of $^{28}$Mg all show a similar
dependence as the proton separation energy $S_\text{p}$ is allowed to change by a
factor of 40, from 0.5 to 20 MeV, in Fig.~\ref{fig:calc_percent}. Thus, the overall 
percentage of diffraction events has little sensitivity to the details of the shell-model
calculations and their spectroscopic factors. As is
expected the elastic breakup fraction decreases with increasing separation energy
but at nothing like the rate expected were the dependences as $1/\sqrt{S_\text{p}}$,
that would suggest a reduction of more than a factor of 6. This is confirmation,
see e.g. Fig. 3 of \cite{gade08}, that the reaction is not asymptotic. We consider
the physically dominant $1d_{5/2}$ case. Here, the elastic fraction falls by less
than a factor of two, from 28~\% to 17~\%, over the range of separation energies
and these changes correlate much more closely with the weaker sensitivity of the
rms radius, $r_\text{sp}$, of the orbital to $S_\text{p}$ (that ranges from 4.20 to
3.22 fm). For the calculations shown in Fig.~\ref{fig:calc_percent} the $1d_{5/2}$
elastic breakup fraction is found to scale as $r_\text{sp}^{3/2}$ to better than
5\% over the range of $S_\text{p}$ considered. It would be of interest to examine
this dependence further, experimentally, with data on nearby $sd$-shell nuclei
having a range of $S_\text{p}$.

\subsection{Angular acceptance correction\label{cdcc}}

Given the expected and calculated dominance of the ground-state to
ground-state $5/2^+$ transition, a continuum discretized coupled
channels (CDCC) calculation for the elastic breakup from a $1d_{5/2}$
proton orbit is used to estimate those breakup events that are
unobserved due to the angular acceptance of HiRA; as were used for the
lighter projectile data~\cite{bazin09}. The CDCC calculations are
performed using the direct reactions code {\sc fresco}~\cite{fresco}.
The calculations describe $^{28}$Mg in terms of its dominant $^{27}$Na
plus proton ground-state configuration and include $^{27}$Na+p breakup
states for configurations with relative orbital angular momenta
$\ell=0-4$ and relative energy up to 30 MeV. For each  breakup partial
wave this continuum energy interval is divided into 15 bins, while the
coupled channels calculations include projectile-target partial waves
up to total angular momentum 300 and use a matching radius of 50 fm.
The theoretical optical potentials, bound state radial wave functions
and other inputs used were in common with those used in the eikonal
calculations of Section~\ref{eik_calcs}. The model space required for a
stable calculation is far more demanding than in the weakly-bound,
light projectile cases and it was not clear that the calculation and
the derived observables were fully converged. Thus, the correction
should be interpreted as indicative and not fully quantitative.

Within this CDCC model space, the exclusive laboratory frame elastic breakup
differential cross section ($d^3 \sigma / dE_p d\Omega_p d\Omega_r$), see e.g.
Ref. \cite{tostevin01}, was computed. This differential cross section, fully-integrated
over the proton and residue final state variables was then compared with that
integrated over the experimental acceptances; that the $^{27}$Na residue travels
in the forward direction with $\Delta \Omega_r$= 21 msr and that $\vartheta_p =
9 - 56^\circ$; see also \cite{bazin09} for further details. Based on this three-body
CDCC model analysis we estimate that 15\% of the elastic breakup events will be
unobserved by the detector (polar angle) acceptances of the present experimental
setup. This provides an approximate angular acceptance correction
factor to the measured (lower-limit) elastic breakup fractions. If applied to the
fraction measured in focused mode (12.5(17)~\% presented in Section
\ref{exp_anal}), which is better suited since the angles of protons
can be determined with high accuracy, a deduced elastic breakup fraction of
14.7(20)\% is obtained. This experimental value is shown in
Fig.~\ref{fig:exp_percent}. If instead the value measured in the
dispersion matched mode is corrected for the proton angle acceptance,
the resulting elastic breakup fraction amounts to 13.2(14)\%.

\subsection{Bound state transition corrections\label{projoff}}

The eikonal model calculations presented above follow the formalism of Ref.
\cite{tostevin01b}, specifically, Eqs. (2) and (4--8), and use potential and
size parameters as described in Section \ref{eik_calcs}. These calculations
take only the $i=0$ term in the calculation of the elastic breakup part of the
single-particle cross section in Eq. (6) of \cite{tostevin01b}. The expression
for the elastic breakup cross section to a given residue final state $c$, that
is assumed to be a spectator in the reaction, is written as the following
integral over the projectile center-of mass impact parameters $\vec{b}$,
\begin{eqnarray*}
\sigma_\text{sp}^\text{elas}(c)=\frac{1}{2I+1}\int d\vec{b} \,
\Big[\sum_{M} \langle \phi^{c}_{IM} |\;|{\cal S}_{\rm c}{\cal
S}_{\rm p} |^2|\phi^{c}_{IM} \rangle- \nonumber \\ \sum_{i\gamma m',M}
|\langle \phi^{\,cj'}_{\gamma m'}(i) |{\cal S}_{\rm c} {\cal S}_{\rm p}
|\phi^{c}_{IM} \rangle|^2 \Big], \label{two}
\end{eqnarray*}
where ${\cal S}_{\rm c}$ and ${\cal S}_{\rm p}$ are the residue- and
proton-target elastic scattering S-matrices and $\gamma$ denotes the
core spin substate. Here, $\phi^{c}_{IM}$ is the (normalized) angular
momentum coupled configuration of the core state and the proton in the
projectile ground state, so, for $^{28}$Mg($I^\pi$=$0^+$), $\phi^{c}_{00}
\equiv [c\otimes n(\ell s)j]_{00}$. The $\phi^{\,cj'}_{\gamma m'}(i)$
represent product states of the particular residue (spectator core)
state $c\gamma$ and a proton in a configuration $[n' (\ell' s) j'm']$.
The $i=0$ term represents the $n' \ell'_{j'} \equiv 1d_{5/2}$, $1d_{3/2}$ 
or $2s_{1/2}$ proton bound state, respectively, for the $5/2^+$, $3/2^+$ 
and $1/2^+$ residue final states of importance here, the same $n \ell_{j}$ 
orbital as appears in $\phi^{c}_{00}$. Extra terms, with $i\neq 0$, 
represent other possible bound proton single-particle configurations 
with respect to the core state $c$. So, the elastic breakup calculations 
presented in Table \ref{tab:tab1},
truncated to $i=0$, assume that all projectile-target interactions that
remove the proton from the given entrance channel configuration, $1d_{5/2}$,
$1d_{3/2}$ or $2s_{1/2}$, lead to continuum states of $^{27}$Na and the
proton and elastic breakup. These neglect proton single-particle transitions
to other bound states of the proton and the residue $c$. As is clear from
the structure of the equation for $\sigma_\text{sp}^\text{elas}(c)$,
the inclusion of such transitions will necessarily reduce the calculated
elastic breakup cross section.

The role of such transitions between bound configurations was calculated
and found to have a very minor effect on the single particle cross sections
(stripping plus diffraction) for light neutron-rich nuclei (of order $3\%$
in Ref. \cite{tostevin01b}). This was due in part to the importance and
dominance of the stripping mechanism, even for weakly-bound nucleons,
seen in Fig.~\ref{fig:calc_percent}, and to the likelihood of a small
number of such bound $i\neq 0$ configurations when removing a weakly-bound
nucleon of an excess species. Such terms have since been neglected in
comparisons with data.

Since the present work involves a direct and exclusive measurement of
the elastic breakup contribution and a well-bound proton, it is opportune
to estimate this effect in this case. In addition to the $i=0$ terms,
the set of possible bound configurations ($i=1,2,3$) may involve proton
$1d_{5/2}$, $1d_{3/2}$, $2s_{1/2}$ and $1f_{7/2}$ configurations. Whether
there is bound $1f_{7/2}$ proton strength with respect to the core states
is unclear. A spherical HF calculation for $^{28}$Mg using the Skyrme SkX
interaction \cite{SkX} does bind the $\pi 1f_{7/2}$ orbital by $\approx $1
MeV. The reduced radius parameter $r_0$ consistent with HF constraints is
1.209 fm for this $7/2^-$ orbital.

The eikonal calculations of $\sigma_\text{sp}^\text{elas}(c)$ have been
repeated. In addition to the $i=0$ calculations of Table \ref{tab:tab1},
reproduced and denoted 0 in Table \ref{tab:tab2}, we show there the results
for calculations that include bound transitions assuming a $1d_{5/2}$,
$1d_{3/2}$ and $2s_{1/2}$ space (denoted $ds$) and also when including an
assumed $1f_{7/2}$ bound state at the calculated HF separation energy of
0.84 MeV (denoted $dsf$). Table \ref{tab:tab2} shows, in each case, the
theoretical partial and inclusive removal cross sections, calculated as
in Table \ref{tab:tab1}, and the computed percentage contribution from
elastic breakup. As must be the case, this elastic breakup cross section
fraction is reduced as the space of possible proton transitions to bound
states is increased, since these transitions remove flux hitherto assumed
to lead to breakup. However, as was found in \cite{tostevin01b}, the effect
on the inclusive proton removal cross section is small ($\lesssim 3\%$)
and within the errors for typical data sets involving well-bound nucleon
removal.

\begin{table}[ht]
\caption{\label{tab:tab2} Theoretical cross sections for one-proton
removal from $^{28}$Mg at 93 MeV/nucleon on a $^9$Be target. Calculations
use the USD shell model spectroscopic factors of Table \ref{tab:tab1}. The
predicted percentages of elastic breakup events are shown in the columns
headed elas. The three sets of calculations, labeled $0$, $ds$ and
$dsf$, result from eikonal model calculations that neglect (case 0) and
include (cases $ds$ and $dsf$) the effects of proton single-particle
transitions between bound states upon the elastic breakup calculation.
Full details are given in the text of Section \ref{projoff}.}
\begin{tabular}{cc|cc|cc|ccc}
\hline \hline
\ \ $J^\pi$\ \ &$E_{\rm x}$&\ $\sigma_{\rm th}$\ \ & \ elas\ \ &\ $
\sigma_{\rm th}$\ \ & \ elas\ \ &\ $\sigma_{\rm th}$\ \ & \ elas\ \ \\
        &(MeV)  &(mb)&(\%)&(mb)&(\%)&(mb)&(\%)\\
        &       &$0$&$0$&$ds$&$ds$&$dsf$&$dsf$\\
\hline
 $5/2^+$    &0.000&47.22& 17.9&46.14& 16.0&45.71& 15.2\\
 $3/2^+$    &0.014& 3.59& 17.8& 3.51& 16.0& 3.50& 15.6\\
 $1/2^+$    &1.630& 4.78& 19.3& 4.60& 16.0& 4.52& 14.6\\
\hline
Sum &            &55.59& 18.0&54.25& 16.0&53.73& 15.2\\
\hline\hline
\end{tabular}
\end{table}

\begin{figure}[ht]
\centering
\includegraphics[width=\columnwidth]{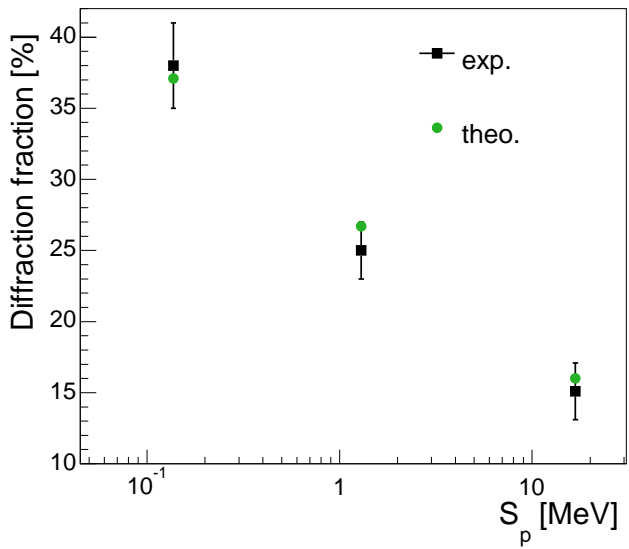}
\caption{(Color online) Experimental (black squares) and theoretical (green circles) fractional contribution (\%) of the elastic breakup
(diffraction) reaction mechanism to the proton removal cross section as a function
of the proton separation energy. In addition to the present study the
figure also shows the previously
measured cases of weakly-bound proton removals from $^9$C and $^8$B~\cite{bazin09}.}
\label{fig:exp_percent}
\end{figure}

The calculations suggest, however, that the eikonal model percentage
of elastic breakup events shown in Table \ref{tab:tab1} (i.e.
calculation 0) and in Fig.~\ref{fig:calc_percent}, of 18\%, is too
large. The most reliable estimate based on these extended calculations,
summarized in Table \ref{tab:tab2}, is that the elastic fraction is
16\%, since the additional assumption of bound $\pi 1f_{7/2}$ states
is speculative. Fig.~\ref{fig:exp_percent} shows that these theoretical values are in good agreement with
that, 14.7(20)\%, deduced from the present more exclusive measurement.
For completeness, Fig.~\ref{fig:calc_percent} also shows the
calculated and experimental values
for the earlier-studied cases, of proton removal from $^9$C and $^8$B
\cite{bazin09}. For all three cases, spanning a wide range in
separation energies, theoretical calculations agree with the
experimentally determined fractional contribution of the elastic
breakup reaction mechanism.

\section{Summary and conclusions}

In summary, we have studied the one-proton removal reaction from the neutron-rich
$sd$-shell nucleus $^{28}$Mg at the intermediate energy of 93 MeV per nucleon. Light
charged particles were detected in coincidence with the fast $^{27}$Na residues
allowing the missing mass spectrum to be measured. The distribution of events in the
missing mass spectrum allowed the determination of the relative contribution of
elastic breakup to the removal reaction in this case of a well-bound proton.
This deduced fraction of elastic breakup events was found to be in good agreement
with the eikonal model calculations of the reaction yields within the experimental
and theoretical uncertainties, extending the earlier-reported agreement for
removal reactions involving weakly-bound protons.

\acknowledgments
This work was supported by the National Science Foundation under Grant No. PHY-0606007
and PHY-0757678 and by DOE/NNSA (National Nuclear Security Administration) Grant No.
DE-FG55-08NA28552. JAT acknowledges the support of the United Kingdom Science and
Technology Facilities Council (STFC) under Grants Nos. ST/J000051/1 and ST/L005743/1.

\newpage
\bibliography{breakup}

\end{document}